# Structure determination of flat honeycomb Bi grown on Ag(111)

Ziyong Zhang[a], Xiaobin Chen[a], and Takeshi Nakagawa[a,b,*]

[a] Science and Engineering of Materials and Devices, Interdisciplinary Graduate School of Engineering Sciences, Kyushu University, 6-1 Kasuga Koen, Kasuga, Fukuoka, 816-8580, Japan.
[b] Department of Advanced Materials Science and Engineering, Faculty of Engineering Sciences, Kyushu University, 6-1 Kasuga Koen, Kasuga, Fukuoka, 816-8580, Japan.

**ABSTRACT**. Honeycomb bismuthene structures on Ag(111) were investigated using low-energy electron diffraction (LEED) and density functional theory. LEED $I(V)$ analysis revealed that 0.5 monolayer (ML) of Bi forms an ultraflat honeycomb lattice with negligible buckling at ~120 K, which transforms into other structures upon warming to room temperature. A similar flat bismuthene structure also forms in Mn/Bi/Ag(111), which remains stable even at room temperature. Mn deposition on (p × $\sqrt{3}$)-rect Bi/Ag(111) induces Bi surface segregation, as confirmed by X-ray photoelectron spectroscopy, resulting in a p(2 × 2) honeycomb bismuthene. The detailed structural investigation provides fundamental insights into the characterization of two-dimensional topological properties of bismuthene grown on Ag(111).

## I. INTRODUCTION.

Two-dimensional topological insulators (2D TIs), which exhibit large bulk gaps and host one-dimensional topological edge states, have been considered promising candidates for achieving dissipationless transport and realizing quantum spintronic devices [1–4]. Two-dimensional topological materials composed of group-IV elements, which are represented by graphene, silicene, germanene, and stanene, have been discovered and extensively studied. However, most of these materials exhibit relatively small energy gaps, requiring low-temperature conditions and thereby limiting their practical applications [5–8]. Therefore, it is believed that materials composed of other elements are required for applications at room temperature.

Over the last decade, Group-V elemental 2D materials have attracted increasing attention because of their intriguing properties, such as tunable band gap, high carrier mobility, and promising transition to a topological nontrivial phase [9]. Bi, which is a heavy element, exhibits strong spin-orbit coupling and is considered a two-dimensional topological insulator with a large nontrivial topological gap. In particular, it has been reported to possess a large bulk bandgap, making it a promising candidate for realizing topological properties at room temperature [10].

Bi has a rhombohedral crystal structure. Monolayer Bi, which is called bismuthene, primarily exhibits two phases, α and β, which crystallize in the (110) and (111) orientations, respectively. The α-bismuthene (α-Bi) has a puckered black-phosphorus-like structure with a rectangular unit cell and a paired bilayer (BL) structure [11]. On the other hand, β-bismuthene (β-Bi) exhibits a hexagonal (111) structure (A7 crystal structure), and it also grows in a bilayer arrangement. Three atoms located in the same plane are horizontally shifted by 60 degrees, positioning themselves alternately at upper and lower sites to form a buckled honeycomb structure [12,13]. β-Bi has been confirmed to be a 2D TI with highly localized topological edge states, as demonstrated theoretically [14,15] and experimentally for a single Bi(111) BL on $Bi_2Te_3$ substrate [16,17]. Alternatively, a planar honeycomb form of bismuthene, theoretically predicted to be a 2D topological insulator, has been synthesized on SiC(0001) [18]. Scanning tunneling spectroscopy measurements revealed a metallic edge state and a large topological energy gap of ~0.8 eV, which is attributed to the substrate-induced tensile strain.

Two types of honeycomb bismuthene structures on Ag(111) have recently been reported: one formed by Bi deposition at low temperatures [19], and the other by subsequent deposition of Mn and Bi at room temperature [20]. Scanning tunneling microscopy (STM) observations indicate flat honeycomb structures with a p(2 × 2) superstructure in both cases. Angle-resolved photoemission spectroscopy (ARPES) measurements on the Bi/Ag(111) showed good agreement with band calculations for ultraflat bismuthene [9], featuring a topological gap of ~1.4 eV at the $\overline{K}$ point due to Bi $p_{xy}$ orbitals. These results suggest the formation of flat bismuthene, although quantitative structural determination has yet to be performed.

Here, we report the atomic structure of these two honeycomb bismuthene using dynamical low-energy electron diffraction (LEED) analysis. To validate the LEED-derived structures, we conducted density functional theory (DFT) calculations, which showed good agreement. Since the Mn/Bi/Ag(111) may involve a segregation process, we applied X-ray Photoelectron Spectroscopy (XPS) to investigate this phenomenon and successfully provided evidence for its occurrence during the Mn deposition.

## II. EXPERIMENTAL AND CALCULATIONS

Experiments were performed in an ultra-high vacuum chamber with a base pressure of $8 \times 10^{-9}$ Pa, equipped with a four-grid LEED and hemispherical analyzer (PHOIBOS100) for X-ray photoelectron spectroscopy (XPS). XPS was conducted using a non-monochromatic Mg K$\alpha$ X-ray source with a photon energy of 1253.6 eV. Bi and Mn were deposited on Ag(111) using Ta crucibles. The purities of the Bi and Mn sources were 99.9999% and 99.97%, respectively. Ag(111) was cleaned with repeated cycles of Ar$^+$ ion sputtering (1 kV, 2 μA, 15 min) and subsequent annealing at 400 K for 30 minutes, until a sharp $(1 \times 1)$ LEED pattern was observed. For the Bi/Ag(111), deposition was performed at a low temperature of 120 K to prevent alloy formation [21]. In contrast, the Mn/Bi/Ag(111) was prepared entirely at room temperature.

The LEED patterns and $I(V)$ curves were recorded at 120 K [22] using a CCD camera. We used the Barbieri/Van Hove symmetrized automated-tensor LEED package to calculate theoretical $I(V)$ curves [23]. We performed the Pendry reliability factor ($R_p$) to evaluate the agreement between the experimental curves and theoretical ones [24]. The error range of the structural parameters were obtained from the variance of the $R_p$ using the formula $\Delta R_p = R_{min} (8V_{0i}/E_t)^{1/2}$, where $R_{min}$ is the minimum of $R_p$, $E_t$ is the total energy width used for the $I(V)$ analysis, and $V_{0i}$ is the imaginary part of the inner potential, respectively. $V_{0i}$ was fixed at - 5 eV. The total energy width for Bi/Ag(111) and Mn/Bi/Ag(111) were 3331 eV, and 4299 eV, respectively. All the crystal structures were visualized using VESTA [25]. The lattice constant of Ag used for LEED $I(V)$ analysis was 4.09 Å.

To support the LEED results, we performed DFT calculations using the Vienna *ab initio* simulation package (VASP) [26,27], employing the projector augmented-wave (PAW) potentials [28]. The exchange-correlation function was approximated within the generalized gradient approximation (GGA) as parametrized by Perdew, Burke, and Ernzerhof (PBE) [29]. The energy cutoff for plane waves was 500 eV. A Monkhorst-Pack k-point mesh of $9 \times 9 \times 1$ was used for surface relaxation [30]. The surface slabs were modeled with 6 layers of Ag and a vacuum region of approximately 20 Å. The lattice constant of Ag was set to 4.15 Å, based on a bulk cell optimization. Structural optimization was carried out until the forces were less than $10^{-3}$ eV/Å. Bader charge analysis [31] was performed to investigate the charge transfer and to evaluate the stability of bismuthene.

## III. RESULTS
### A. Bi/Ag(111) at low temperature

A p$(2 \times 2)$ bismuthene structure was prepared by depositing Bi on Ag(111) at approximately 120 K, following the procedure described in [19]. Figure 1(a) displays the corresponding p$(2 \times 2)$ LEED pattern. We performed structural analysis for the bismuthene model proposed in [19]. The resulting $I(V)$ analysis shows good agreement between the experimental and calculated curves, as shown in Fig. 1(b), yielding a Pendry R-factor of 0.20.

A detailed structural analysis was performed for the bismuthene model, and the structural parameters and their corresponding uncertainties, obtained from LEED $I(V)$ analysis and DFT calculations, are summarized in Table SI. The structural parameters by the LEED $I(V)$ and DFT are in good agreement. The optimized Debye temperature of Bi is 96 K, which is consistent with the surface Debye temperature of the first two atomic layers on Bi(111) [13]. The optimized structure is shown in Fig. 1(c). The buckling of the bismuthene on Ag(111) is only 0.02 Å, resulting in an ultraflat honeycomb structure, consistent with previous STM and DFT studies [19]. In the $\beta$-phase Bi, a buckled honeycomb structure is formed, similar to that of phosphorene [12,13]. The observed buckling in Bi/Ag(111) is negligible compared to that in β-phase Bi by LEED $I(V)$, freestanding bismuthene and bulk Bi by DFT calculation, which exhibits a buckling of 1.60 Å, 1.73 Å, 1.65 Å, respectively [13,32,33]. The interlayer distance between bismuthene and the Ag substrate is 2.44 Å (2.49Å: DFT), which is smaller than the 2.75 Å for bismuthene on SiC(0001) [18], or the height of 2.66 Å for Bi atoms on Ag(111) based on a ball model. This suggests a strong interaction between bismuthene and Ag(111), which suppresses buckling .

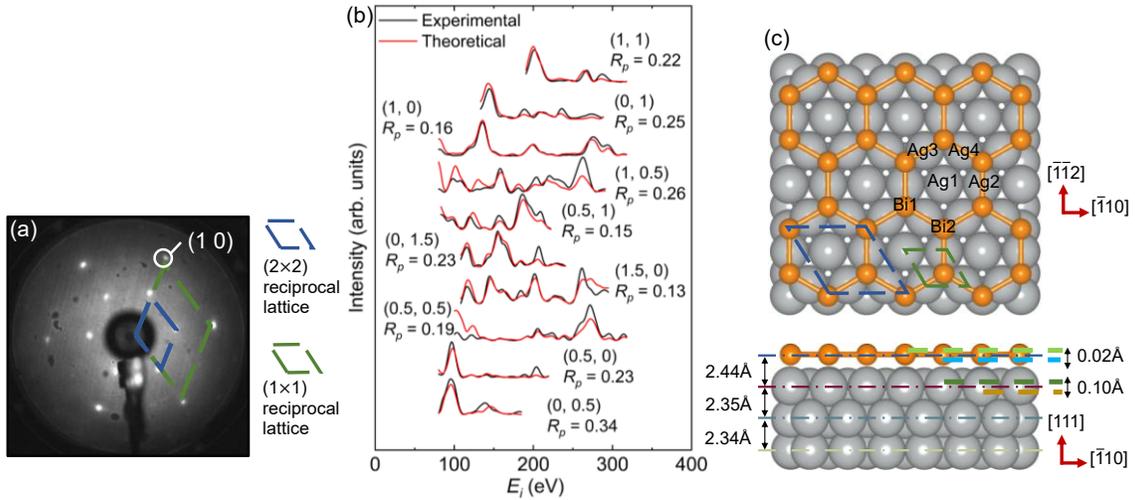

FIG 1. Structural results for Ag(111)-p$(2 \times 2)$-Bi. (a) LEED pattern with an incident electron energy of 99 eV. (b) Comparison of experimental and theoretical LEED $I(V)$ curves. R-factor for each beam is shown below the beam index. The total R-factor is 0.204. (c) Top and side views of the best-fit structure model.

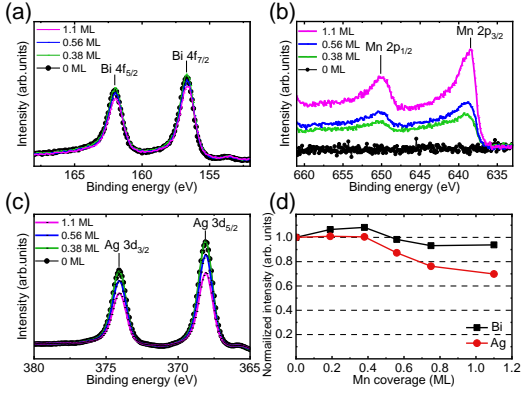

FIG 2. Photoelectron spectra of (a) Bi 4f, (b) Mn 2p, and (c) Ag 3d with increasing Mn coverage on Ag(111)-(p × √3)-Bi. (d) Normalized peak intensities of Bi 4f and Ag 3d as a function of Mn coverage.

The interaction between Ag and Bi also modifies the structure of the first Ag layer. LEED results indicate that the Ag atom (Ag1) located beneath the center of the honeycomb shifts upwards by 0.12 Å relative to the surrounding Ag atoms. This strong interaction drives the Bi atoms to seek higher coordination with the Ag substrate. Consequently, we propose that bismuthene becomes more stable when the height of Ag1 increased. However, DFT calculations suggest that Ag1 sinks by 0.03 Å relative to the surrounding Ag atoms. This discrepancy indicates that further investigation is necessary.

### B. Mn/Bi/Ag(111) at room temperature

The p(2 × 2) honeycomb bismuthene on Ag(111) transforms into a Bi-Ag surface alloy or (p × √3)-rect Bi overlayer upon increasing the temperature from 120 K to room temperature. Recently, a distinct honeycomb structure was observed at room temperature by depositing Mn onto Bi/Ag(111) [20]. The (p × √3)-rect Bi structure with $\theta_{Bi}$ ~ 0.75 ML transforms into a p(2 × 2) structure upon Mn deposition. To gain insight into the positions of Mn and Bi atoms in Mn/Bi/Ag(111), we performed XPS measurements during Mn deposition up to 1.1 ML, as shown in Fig. 2. With increasing Mn coverage, the intensity of the Mn 2p core-level peak increases gradually [Fig. 2(b)]. Figure 2(d) shows the variation of Ag 3d and Bi 4f peak intensities as a function of Mn coverage. As Mn deposition increases, the Ag 3d intensity decreases significantly, while the Bi 4f intensity shows only a slight reduction. A minor increase in Bi 4f intensity between 0.2 and 0.4 ML may be attributed to changes in backscattering effects [34]. These XPS trends suggest that Bi atoms remain in the overlayer even after Mn deposition. The gradual decrease in Bi 4f intensity above 0.4 ML can be attributed to two possible mechanisms: (i) a structural transition of Bi from a buckled to an ultraflat configuration, or (ii) partial alloying between Bi and Mn atoms.

We performed LEED $I(V)$ analysis for Bi/Mn/Ag(111). Figures 3(a) and 3(b) show the (p × √3)-rect Bi LEED pattern with $\theta_{Bi}$ ~ 0.75 ML and the p(2 × 2) pattern observed after Mn deposition, respectively. We analyzed 16 structural models with p3m1 symmetry, as detailed in the Supplemental Material (Fig. S1), considering Bi coverage ranging from 0.25 to 0.75 ML and restricting Mn atoms to the top two atomic layers. The analysis began with models in which Bi atoms segregate to form a honeycomb structure on Mn/Ag(111), as shown in Figs. S1(a) and S1(b). Mn atoms were assumed to occupy either fcc or hcp sites, based on previous reports that Mn can form both fcc and hcp monolayers on Ag(111) [35]. However, neither model reproduced the experimental $I(V)$ curves satisfactorily, yielding relatively high $R_p$ of 0.324 for the fcc model and 0.621 for the hcp model.

We then conducted structural analysis of Bi/Mn and Ag/Mn alloy models in the second layer, keeping the Bi honeycomb on the surface. Among these, the structure shown in Fig. 3(d) yielded the lowest $R_p$ of 0.161, indicating good agreement between experimental and calculated I(V) curves [Fig. 3(c)]. The optimized structure features a Bi honeycomb structure on a Mn$_3$Bi alloy in the second layer with Mn and Ag occupying fcc sites.

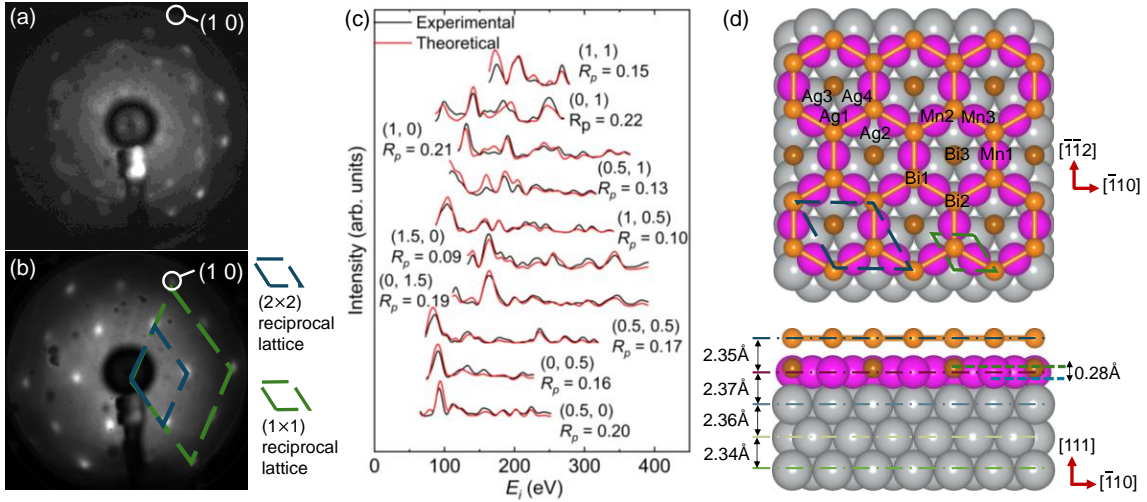

FIG 3. (a) LEED pattern of Ag(111)-(p × √3)-Bi at ~1 ML. (b) (2 × 2) LEED pattern after subsequent Mn deposition. The incident electron energies are 43 eV and 71 eV for (a) and (b), respectively. (c) Comparison of experimental and theoretical LEED I(V) curves for Mn/Bi/Ag(111). R-factor for each beam is shown below the beam index. The total R-factor is 0.161. (d) Top and side views of the best-fit structure model.

Table SII summarizes the structural parameters and their uncertainties obtained from LEED I(V) analysis and DFT calculations, showing overall consistency. The optimized Debye temperature of Bi was found to be reduced to 90 K, consistent with the structural flattening discussed earlier. Mn/Bi/Ag(111) forms an ultraflat honeycomb Bi layer with negligible buckling (< 0.01Å), in agreement with STM observations [20] and similar to the Bi/Ag(111) structure. The interlayer spacing between the Bi layer and the Mn$_3$Bi layer is 2.35 Å, which is smaller than that for Bi on Ag(111) (2.44 Å), suggesting a stronger interaction between the Bi layer and the Mn$_3$Bi alloy.

Based on these structural results, we propose that Bi segregation occurs during Mn deposition, as illustrated in Fig. 4(a). Some Bi atoms segregate to the surface and form an ultraflat bismuthene layer, while others form the Mn$_3$Bi alloy in the second layer. This mechanism explains the transformation from (p × √3)-rect structure with $\theta_{Bi}$ ~ 0.75 ML to the (2 × 2) honeycomb structure, corresponding to an ideal first layer Bi coverage of 0.5 ML and 0.25 ML in the Mn$_3$Bi layer. This is consistent with the slight reduction observed in the Bi 4f intensity in the XPS spectra, as shown in Fig. 2(d).

The flat honeycomb on Mn$_3$Bi layer can be attributed to two factors: Bi-Mn alloy formation and the intrinsic tendency of Bi to adopt a buckled Bi(111)-like configuration. The Bi–Mn bond lengths in the final structure were 2.93 Å (Bi1–Mn), 2.95 Å (Bi2–Mn), and 2.90 Å (Bi3–Mn), in good agreement with the 2.916 Å Bi–Mn bond length in MnBi [36]. Additionally, the Bi–Bi bond lengths between the flat honeycomb Bi atoms (Bi1 and Bi2) and the subsurface Bi atom (Bi3) are 3.96 Å, closely matching the 3.93 Å Bi–Bi bond length in MnBi. Structurally, the Bi atoms form a flat monolayer honeycomb with an additional Bi atom located beneath the center of each hexagon. This arrangement resembles the buckled honeycomb structure of the Bi(111) surface ($\beta$-Bi). Thus, the structure shown in Fig. 3(d) reflects both the influence of Bi–Mn alloy formation and the intrinsic Bi(111)-like configuration. This interpretation is further supported by the lack of significant chemical shifts in the Bi core-level peak observed in Fig. 2(a).

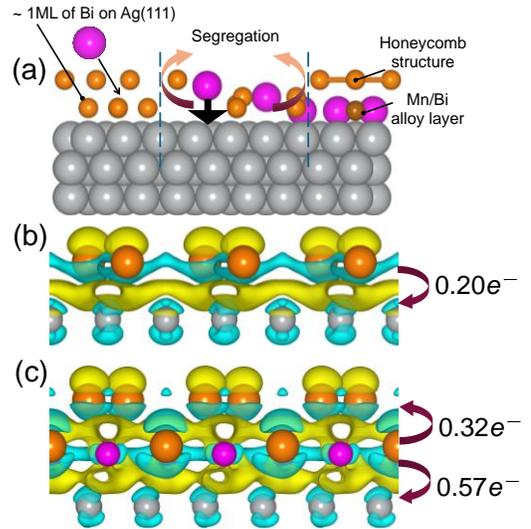

FIG 4. (a) Schematic illustration of the segregation process induced by Mn deposition on Ag(111)-(p × √3)-Bi overlayer structure. (b), (c) Charge density difference of Bi/Ag(111) and Mn/Bi/Ag(111), respectively, based on Bader charge analysis (see Table SIII and SIV). The amount of electron transfer shown in (b) and (c) are given per unit cell.

Honeycomb structures have been observed on various metal surfaces, such as Li/Cu(111) [37], Sn/Ag(111) [38], and Sb/Ag(111) [39], where surface alloying stabilizes the honeycomb lattices. In these systems, atoms in the second alloy layer may play a crucial role in stabilizing the low-coordination honeycomb networks, as observed in the case of Bi in $Mn_3Bi$ alloy on Ag(111).

Bader charge analysis was performed for both Bi/Ag(111) and Mn/Bi/Ag(111) to examine charge transfer between bismuthene and the substrate. Detailed results are provided in the Supplemental Material (Tables SIII and SIV). In Bi/Ag(111), each Bi atom in bismuthene transfers approximately 0.1 e⁻ to the Ag substrate, as illustrated in Fig.4(b). Conversely, in Mn/Bi/Ag(111), Mn atoms are located between the bismuthene layer and the Ag substrate. Due to the lower electronegativity of Mn, each Bi atom in bismuthene gains an average of 0.16 e⁻ from the underlying layers as illustrated in Fig.4(c). This enhanced charge transfer contributes to the improved stability of bismuthene at room temperature [40]. Furthermore, we suggest that the elements similar to Mn, such as Cr, may facilitate the formation of bismuthene through analogous segregation mechanism.

## V. CONCLUSIONS

In conclusion, we determined the atomic structures of two types of bismuthene grown on the Ag(111) substrate using LEED $I(V)$ analysis and DFT calculations. The bismuthene grown onto Ag(111), which is stable only at low temperature, exhibits an ultraflat geometry with negligible buckling (< 0.02 Å). The other bismuthene emerged via a segregation process following Mn deposition onto approximately 0.75 ML of Bi on Ag(111) at room temperature, and is similarly flat, with negligible buckling (~0.01 Å). In the Mn/Bi/Ag(111), bismuthine does not formed on a Mn layer but rather on a $Mn_3Bi$ surface alloy, where a subsurface Bi atom resides beneath the center of the Bi hexagon. These findings provide a fundamental basis for further investigation of the two-dimensional topological properties of bismuthenes on Ag(111).

## ACKNOWLEDGMENTS

This work was supported by JSPS KAKENHI grants (Grant No. 22K04928). The numerical computation was carried out partially using the computer resource offered under the category of General Project by Research Institute for Information Technology, Kyushu University and the Supercomputer Center, Institute for Solid State Physics, University of Tokyo. We thank Dr. Yuki Fukaya and Dr. Kazutoshi Takahashi for fruitful discussion.

# Supplemental Materials:

## Structure determination of flat honeycomb Bi grown on Ag(111)

Ziyong Zhang[a], Xiaobin Chen[a], and Takeshi Nakagawa[a,b,*]

[a] Science and Engineering of Materials and Devices, Interdisciplinary Graduate School of Engineering Sciences, Kyushu University, 6-1 Kasuga Koen, Kasuga, Fukuoka, 816-8580, Japan.
[b] Department of Advanced Materials Science and Engineering, Faculty of Engineering Sciences, Kyushu University, 6-1 Kasuga Koen, Kasuga, Fukuoka, 816-8580, Japan.

## I. Structure parameters of Bi/Ag(111) and Mn/Bi/Ag(111)

TABLE SI. Atomic coordinates and their uncertainties for the optimized structure of Bi/Ag(111) as determined by LEED *I(V)* analysis. Positions marked with * are symmetry-constrained and fixed. The optimized Debye temperature of Bi and 1st layer Ag are 96 and 147 K, respectively.

| Atoms | LEED | | | | | | DFT | | |
|---|---|---|---|---|---|---|---|---|---|
| | [$\bar{1}$10] | | [$\bar{1}\bar{1}$2] | | [111] | | [$\bar{1}$10] | [$\bar{1}\bar{1}$2] | [111] |
| Bi1 | 0.00 | * | 3.34 | * | 2.47 | ±0.03 | 0.00 | 3.39 | 2.48 |
| Bi2 | 2.89 | * | 1.67 | * | 2.45 | ±0.03 | 2.93 | 1.69 | 2.49 |
| Ag1 | 0.00 | * | 0.00 | * | 0.09 | ±0.04 | 0.00 | 0.00 | -0.03 |
| Ag2 | 2.89 | ±0.05 | -0.01 | ±0.05 | 0.00 | ±0.02 | 2.93 | 0.00 | 0.00 |
| Ag3 | -1.44 | ±0.05 | 2.50 | ±0.05 | 0.00 | ±0.02 | -1.47 | 2.54 | 0.00 |
| Ag4 | 1.44 | ±0.05 | 2.50 | ±0.05 | 0.00 | ±0.02 | 1.47 | 2.54 | 0.00 |

TABLE SII. Atomic coordinates and their uncertainties for the optimized structure of Mn/Bi/Ag(111) as determined by LEED *I(V)* analysis. Positions marked with * are symmetry-constrained and fixed.

| Atoms | LEED | | | | | | DFT | | |
|---|---|---|---|---|---|---|---|---|---|
| | [$\bar{1}$10] | | [$\bar{1}\bar{1}$2] | | [111] | | [$\bar{1}$10] | [$\bar{1}\bar{1}$2] | [111] |
| Bi1 | 0.00 | * | 3.34 | * | 4.72 | ±0.03 | 0.00 | 3.39 | 4.62 |
| Bi2 | 2.89 | * | 1.67 | * | 4.72 | ±0.03 | 2.93 | 1.69 | 4.64 |
| Bi3 | 0.00 | * | 0.00 | * | 2.59 | ±0.05 | 0.00 | 0.00 | 2.64 |
| Mn1 | 2.89 | ±0.05 | -0.02 | ±0.05 | 2.31 | ±0.01 | 2.93 | 0.03 | 2.30 |
| Mn2 | -1.43 | ±0.05 | 2.51 | ±0.05 | 2.31 | ±0.01 | -1.49 | 2.53 | 2.30 |
| Mn3 | 1.43 | ±0.05 | 2.51 | ±0.05 | 2.31 | ±0.01 | 1.49 | 2.53 | 2.30 |
| Ag1 | 0.00 | * | -3.34 | * | 0.01 | ±0.03 | 0.00 | -3.39 | 0.03 |
| Ag2 | 2.89 | ±0.05 | -3.32 | ±0.05 | 0.00 | ±0.02 | 2.93 | -3.37 | 0.00 |
| Ag3 | -1.46 | ±0.05 | -0.84 | ±0.05 | 0.00 | ±0.02 | -1.48 | -0.85 | 0.00 |
| Ag4 | 1.46 | ±0.05 | -0.84 | ±0.05 | 0.00 | ±0.02 | 1.48 | -0.85 | 0.00 |

**II. Examined models for the Ag(111)-(2 × 2)-Bi&Mn**

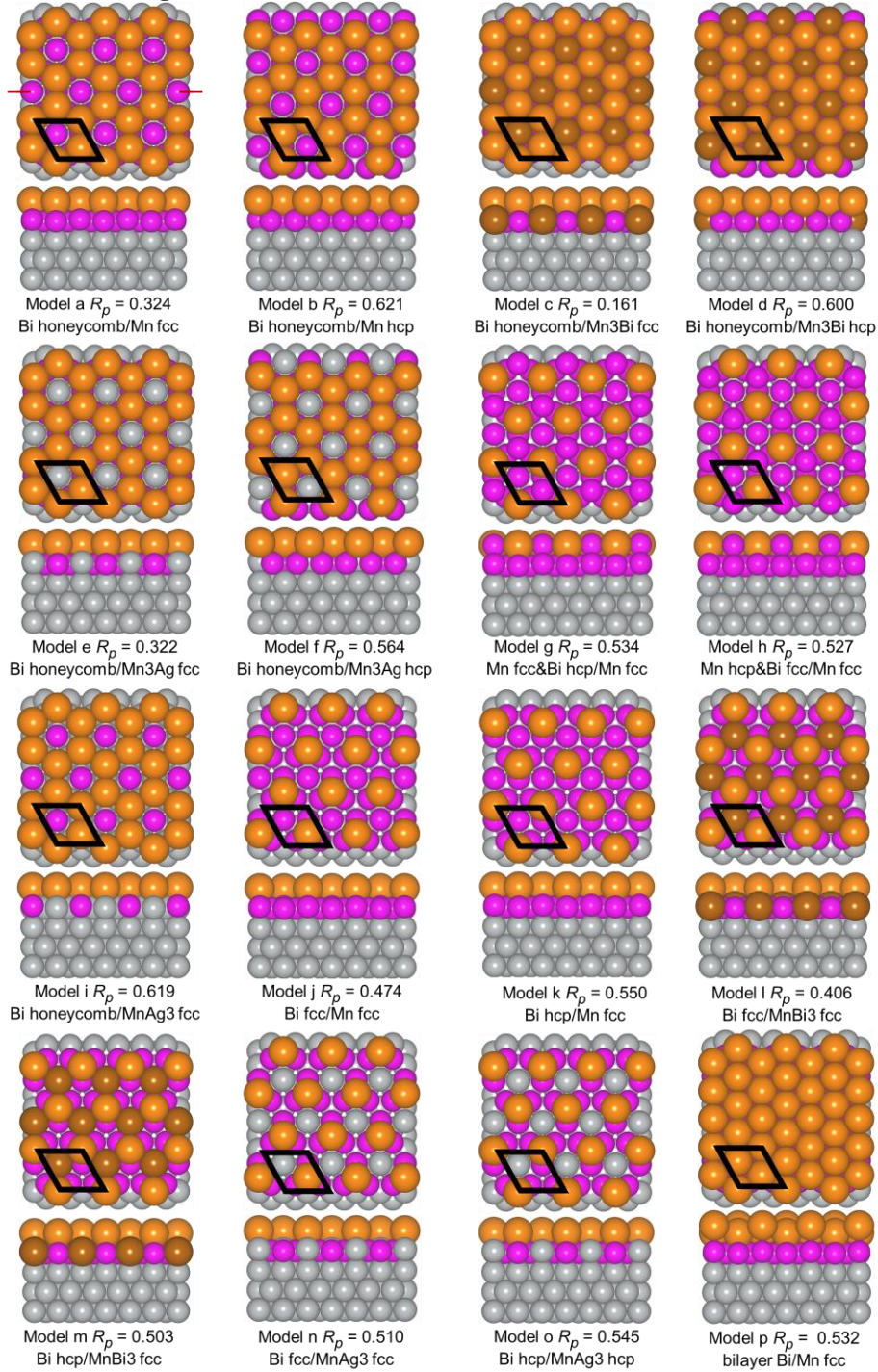

FIG. S1. Ball models of examined 16 structures of Ag(111)-p(2 × 2)-Bi&Mn (top & cross sectional view). Surface atomic structures (the 1st / 2nd layers) are indicated. Unit cells are outlined in black. The red line in Model-a indicates cut-off line for the cross sectional view. The balls in orange represent Bi atoms; the balls in silver indicated Ag atoms.

## III. Formation energy for the Ag(111)-(2 × 2)-Bi&Mn

Although Bi honeycomb/Mn$_3$Bi fcc/Ag(111) (Model c in Fig. S1) showed the lowest R-factor, Bi honeycomb/Mn fcc/Ag(111) (Model a) and Bi honeycomb/Mn$_3$Ag fcc/Ag(111) (Model e) still be concerned as two potential candidates. To confirm the energetic stability of Bi honeycomb/Mn$_3$Bi fcc/Ag(111), we calculated and compared formation energies ($E_{from}$) using DFT. We obtained the formation energy for these three models using the following equations:

$$E_{form} \text{ (Bi honeycomb/Mn)} = E_{total} - E_{Ag(111)} - 2E_{Bi} - 4(E_{Mn(bulk)} - \mu_{Mn})$$

$$E_{form} \text{ (Bi honeycomb/Mn3Bi fcc)} = E_{total} - E_{Ag(111)} - 3E_{Bi} - 3(E_{Mn(bulk)} - \mu_{Mn})$$

$$E_{form} \text{ (Bi honeycomb/Mn3Ag fcc)} = E_{total} - E_{Ag(111)} - E_{Ag(bulk)} - 2E_{Bi} - 3(E_{Mn(bulk)} - \mu_{Mn})$$

, where $E_{total}$ is the total energy for each model, $E_{Ag(111)}$ is the energy of the corresponding (2×2) clean Ag(111) slab with a thickness of 6 ML, $E_{Bi}$ is the energy per atom of bismuth solid with Rhombohedral lattice. We define the formation energy by considering the Mn and Ag atoms at second layer, with $E_{Mn(bulk)}$ and $E_{Ag(bulk)}$, the energy of the bulk Mn and Ag atom, respectively. $\mu_{Mn}$ is defined as the chemical potential difference of Mn relative to its bulk state. We evaluated the formation energy as a function of $\mu_{Mn}$ between +0.5 and −1.0 eV. The formation energies of these three models are shown in Fig.S3. For $\mu_{Mn} < -0.16$ eV, the formation energy of Bi honeycomb/Mn$_3$Bi becomes and is the lowest among the three models.

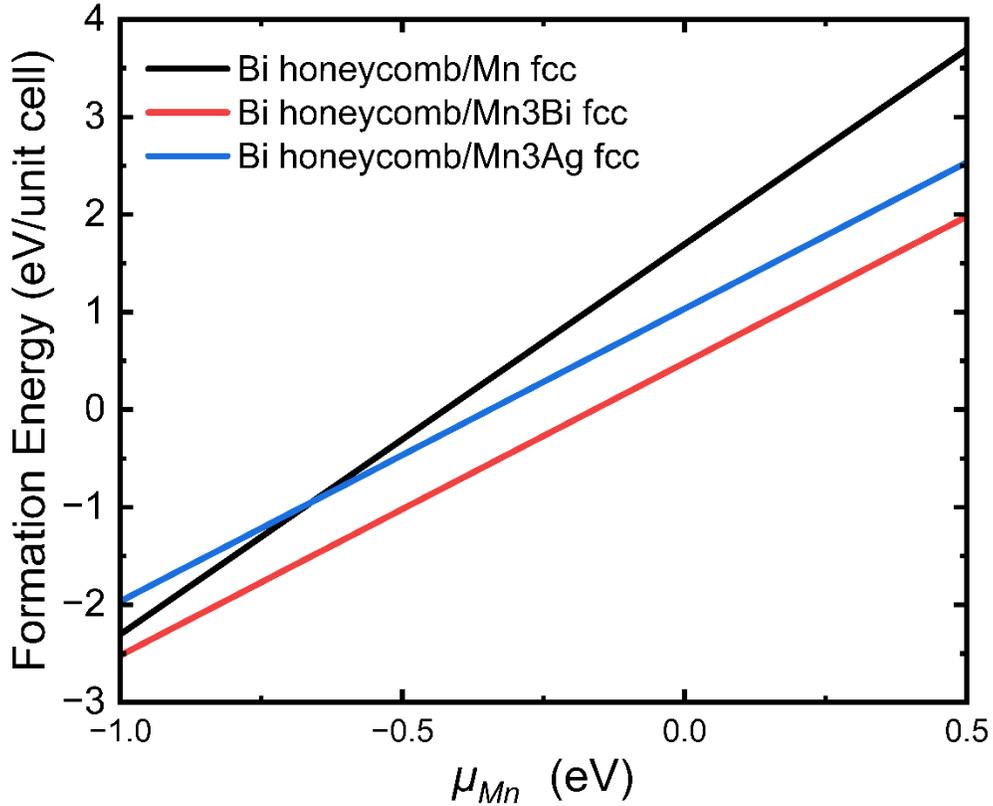

FIG. S2. Variation of formation energy as a function of the chemical potential difference of Mn ($\mu_{Mn}$) for the three models.

**IV. Bader charge analysis and charge density difference for optimized Bi/Ag(111) and Bi/Mn/Ag(111) structures**

Table SIII and SIV shows the calculated charge transfer of Bi/Ag(111) and Bi/Mn/Ag(111), respectively [1].

Table SIII. Charge transfer of Bi/Ag(111). The atoms correspond to those shown in Fig. 1(c).

| Atom | Bader charge transfer ($e^-$) |
|---|---|
| Bi1 | -0.08 |
| Bi2 | -0.12 |
| Ag1 | -0.02 |
| Ag2 | +0.07 |
| Ag3 | +0.08 |
| Ag4 | +0.07 |

Table SIV. Charge transfer of Mn/Bi/Ag(111). The atoms correspond to those shown in Fig. 3(d).

| Atom | Bader charge transfer ($e^-$) |
|---|---|
| Bi1 | +0.22 |
| Bi2 | +0.1 |
| Bi3 | +0.1 |
| Mn1 | -0.34 |
| Mn2 | -0.34 |
| Mn3 | -0.35 |
| Ag1 | +0.21 |
| Ag2 | +0.12 |
| Ag3 | +0.12 |
| Ag4 | +0.12 |

**Reference**
[1] G. Henkelman, A. Arnaldsson, and H. Jónsson, A fast and robust algorithm for Bader decomposition of charge density, Comput Mater Sci **36**, 354 (2006).